   \documentclass[twocolumn,showpacs,preprintnumbers,amsmath,amssymb]{revtex4}
   \usepackage{graphicx,xcolor}    
\usepackage{epstopdf,epsfig}
	\usepackage{bm}
   \usepackage{dcolumn}
   \usepackage{natbib}
	\usepackage{physics}

\begin{document}
   	
   	\title{Ultraviolet absorption spectrum of the half-filled bilayer graphene}
   	
   	\author{V. Apinyan\footnote{Corresponding author. Tel.:  +48 71 3954 284; E-mail address: v.apinyan@int.pan.wroc.pl.}, T. K. Kope\'{c}}
   	\affiliation{Institute for Low Temperature and Structure Research, Polish Academy of Sciences\\
   		PO. Box 1410, 50-950 Wroc\l{}aw 2, Poland \\}
   	
   	\date{\today}

\begin{abstract}
%
We consider the optical properties of the half-filled AB-stacked bilayer graphene with the excitonic pairing and condensation between the layers. Both intra and interlayer local Coulomb interaction effects have been taken into account and the role of the exact Fermi energy has been discussed in details. We have calculated the absorption coefficient, refractive index, dielectric response functions and the electron energy loss spectrum for different interlayer Coulomb interaction regimes and for different temperatures. Considering the full four-band model for the interacting AB bilayer graphene, a good agreement is achieved with other theoretical and experimental works on the subject, in particular, limiting cases of the theory. The calculations, presented here, permit to estimate accurately the effects of excitonic pairing and condensation on the optical properties of the bilayer graphene. The modifications of the plasmon excitation spectrum are discussed in details for a very large interval of the interlayer interaction parameter.        
   	\end{abstract}

   	\pacs{71.35.-y, 71.35.Cc, 72.80.Vp, 74.25.fc, 74.25.N-, 74.25.Gz, 78.67.Wj, 78.20.Ci, 78.40.-q}
   	\maketitle

 \renewcommand\thesection{\arabic{section}}
   	
\section{\label{sec:Section_1} Introduction}
%
The bilayer graphene (BLG), which could be obtained from the mature bulk graphite by the subsequent exfoliation techniques, is recognized as a promising structure in which one can control the induced band gap by the external gate voltage which favorites their applicability to optical processing devices. The external electric field can tune the bilayer graphene from the semimetal to the semiconducting state \cite{cite_1}, by allowing for novel terahertz devices \cite{cite_2} and transistors \cite{cite_3}. The density of charge carriers and the position of the Fermi level, in these systems, can be easily controlled by applying the external gate voltage \cite{cite_4}.
The optical properties in the bilayer graphene system have been studied extensively, both theoretically \cite{cite_5, cite_6, cite_7, cite_8, cite_9, cite_10, cite_11, cite_12, cite_13, cite_14, cite_15, cite_16, cite_17, cite_18, cite_19} and experimentally \cite{cite_20, cite_21, cite_22, cite_23}. The intraband and interband optical transitions in the bilayer graphene have been studied recently in details \cite{cite_16, cite_18, cite_24}. It has been shown that the optical transitions in the BLG system can get substantially modified, after the electrical gating of bilayer graphene \cite{cite_25}, similar to the electrical transport in field-effect transistors \cite{cite_26, cite_27}. A very comprehensive and detailed analysis of the optical transitions, in the bilayer graphene, was given in Ref.\onlinecite{cite_28}, where the comparison of different microscopic models, for the bilayer graphene, is done. The absorption of the electromagnetic field in graphene, without the external magnetic field, was studied in Refs.\onlinecite{cite_7, cite_8, cite_29,cite_30}.

It is well known for the case of usual semiconductors \cite{cite_31} that at the temperatures, where the thermal energy is less than the exciton binding energy the excitons will dominate the optical response and the interaction assisted excitonic transitions will dominate the similar transition between unbound electron-hole pairs. Therefore, one should expect that the excitons will play an equally important role in the electrically gated BLG, being a good example of a quasi two-dimensional semiconductor \cite{cite_32}. 
There exists few works on the excitonic effects on the optical properties of the bilayer graphene \cite{cite_33, cite_34, cite_35, cite_36, cite_37, cite_38, cite_39, cite_40}. It should be particularly distinguished between the effects related to the broadly resonant excitonic states in the optical absorption spectrum and the narrow resonant excitonic states at the high-frequency regime \cite{cite_37}. A very high broadly resonant excitonic red-shift effect has been obtained in Ref.\onlinecite{cite_41} with the help of the ab-initio many-body calculations, but there is no agreement mentioned, with the experimental results, except the case of graphite. The origin of the red-shift is attributed to the background continuum of dipole forbidden transitions. 
  
In addition, recently, a new type of symmetry breaking effects has been found in the AB-stacked bilayer graphene, related to the formation of the excitonic condensate state when tuning the local interlayer Coulomb interaction parameter \cite{cite_42, cite_43, cite_44, cite_45, cite_46}. Particularly, it has been shown that there exists is a Bardeen-Cooper-Schrieffer (BCS)-Bose-Einstein-Condensation (BEC) type crossover in the BLG tuned by the interlayer electron-electron interaction parameter \cite{cite_42, cite_43}. The resulting ``excitonic'' semiconducting state has slightly different band structure that is usually given when applying the electric field. 

In the present paper, we have calculated the optical properties in the bilayer graphene based on the solutions for the excitonic gap parameter, chemical potential and the exact Fermi energy of the BLG, given in a series of our previous works \cite{cite_42, cite_43}. 
We show how the inclusion of the excitonic effects will modify and shift the optical absorption spectrum toward the UV range of the light spectrum. We assume the half-filling condition, for the individual graphene sheets in the bilayer construction, which maintains the charge neutrality for the whole system and, particularly, at each value of the interlayer coupling parameter. As the model for our calculations, we consider the bilayer Hubbard model, which includes the intralayer (in-sheets) $U$-Hubbard Coulomb interaction terms and the local interlayer Coulomb interaction $W$ between the electrons on different sublattices (due to the Bernal stacking structure of the bilayer graphene) and packed vertically at the same lattice space position. The principal advantage of the theory deals with the possibility to consider different coupling regimes in the BLG, which is extremely important for a complete view about the different physical effect, which can take place in the system.  

The paper is organized as follows: in the Section \ref{sec:Section_2}, we introduce the bilayer Hubbard model for the AB-stacked bilayer graphene. In the Section \ref{sec:Section_3}, we calculate the complex optical conductivity function in the BLG with the excitonic pairing and condensation regime. Furthermore, in the Section \ref{sec:Section_4}, we obtain the numerical results for the optical properties in the BLG and we discuss the role of the excitons. Next, in the Section \ref{sec:Section_5}, we give a short conclusion to our paper.  
%
\section{\label{sec:Section_2} The bilayer Hubbard model for BLG}
%
The tight-binding part of the bilayer Hubbard Hamiltonian of our BLG structure is given by 
	\begin{eqnarray}
	H_{0}&=&-\gamma_0\sum_{\left\langle {\bf{r}}{\bf{r}}'\right\rangle}\sum_{\sigma}\left({a}^{\dag}_{\sigma}({\bf{r}})b_{\sigma}({\bf{r}}')e^{-\frac{ie}{\hbar{c}}\int^{{\bf{r}}}_{{\bf{r}}'}{\bf{A}}({\bf{l}})d{\bf{l}}}+h.c.\right)
	\nonumber\\
	&-&\gamma_0\sum_{\left\langle {\bf{r}}{\bf{r}}'\right\rangle}\sum_{\sigma}\left({\tilde{a}}^{\dag}_{\sigma}({\bf{r}})\tilde{b}_{\sigma}({\bf{r}}')e^{-\frac{ie}{\hbar{c}}\int^{{\bf{r}}}_{{\bf{r}}'}{\bf{A}}({\bf{l}})d{\bf{l}}}+h.c.\right)
	\nonumber\\
	&-&\gamma_1\sum_{{\bf{r}}\sigma}\left({{b}}^{\dag}_{\sigma}({\bf{r}})\tilde{a}_{\sigma}({\bf{r}})+h.c.\right)-\sum_{{\bf{r}}\sigma}\sum_{\ell=1,2}\mu_{\ell}n_{\ell\sigma}({\bf{r}}).
	\nonumber\\
	\label{Equation_1}
	\end{eqnarray}
	Here, ${a}^{\dag}_{\sigma}({\bf{r}})$ (${a}_{\sigma}({\bf{r}})$) and ${b}^{\dag}_{\sigma}({\bf{r}})$ (${b}_{\sigma}({\bf{r}})$) are the electron creation (annihilation) operators at the different sublattice positions $A$  and $B$ in the layer 1 of the BLG, while the same operators with the tilde notations are the electron creation (annihilation) operators in the layer 2 of the BLG. The phase factors near the electron operators appear after the Peierls-Onsager substitution \cite{cite_47, cite_48}, which formally includes the interaction with the electric field component $\bf{E}$ of the external electromagnetic field. The first three terms in the Hamiltonian in Eq.(\ref{Equation_1}) form the usual tight binding model for the BLG structure. The parameter $\gamma_0$ is the intralayer hopping parameter (the most realistic value of the parameter $\gamma_0=3$ eV is given in Ref.\onlinecite{cite_49}, where the electronic structure of bilayer graphene has been determined from the infrared spectroscopy). Next, the parameter $\gamma_1$ describes the interlayer hopping amplitude, between the bottom layer 1 sites $B$ and the top layer 2 sites $\tilde{A}$. We have added also the chemical potential terms in the Hamiltonian, with $\mu_{\ell}$ being the chemical potential in the layer with the index $\ell$ ($\ell=1,2$ describes different layers in the BLG). Initially, we suppose that the BLG is in the Grand canonical equilibrium state with $\mu_1=\mu_2\equiv \mu$.  After expanding the tight-binding part, given in Eq.(\ref{Equation_1}) up to the second order in vector potential ${\bf{A}}({\bf{r}})$, we get the following addition to the usual zero field tight-binding term 
	\begin{eqnarray}
	H'_{0}=H_{0}({\bf{A}}=0)-\frac{1}{c}\sum_{\bf{r}}{\bf{A}}({\bf{r}}){\bf{j}}({\bf{r}}).
	\label{Equation_2}
	\end{eqnarray}
	
	The linear term in vector potential is responsible for the paramagnetic part of the total current operator, and we neglect the second order diamagnetic contribution. The interaction term includes the local on-site Hubbard interaction in the individual layers (described by the term-$U$) and between the layers in the BLG (described by the $W$-term).
	
		\begin{eqnarray}
		H_{\rm int}&=&U\sum_{{\bf{r}}}\sum_{\ell\eta}\left[\left(n_{\ell\eta\uparrow}-1/2\right)\left(n_{\ell\eta\downarrow}-1/2\right)-1/4\right]
	\nonumber\\
	&+&W\sum_{{\bf{r}}\sigma\sigma'}\left[\left(n_{1b\sigma}({\bf{r}})-1/2\right)\left(n_{2\tilde{a}\sigma'}({\bf{r}})-1/2\right)-1/4\right].
	\nonumber\\
	\label{Equation_3}
\end{eqnarray}
	
	 The interaction terms $U$ and $W$ in Eq.(\ref{Equation_2}) describe the local Coulomb interactions in the individual layers and between the layers respectively. Namely, the consideration of the local interlayer coupling simplifies the problem considerably when passing into the Fourier space representation of the total Hamiltonian $H_{t}=H_{0}+H_{\rm int}$. 
	Throughout the paper, we will use the parameter $\gamma_0$ as the unit of the energy scale.
 
    %
\section{\label{sec:Section_3} The ac optical conductivity }
    
    In order to calculate the optical absorption, the dielectric response and the refractive index of the BLG with the excitonic pairing interaction we should calculate first the ac optical conductivity in the BLG. For this, there are several ways to treat the bilayer graphene in the presence of the external field. One of those is to assume the presence of the external bias voltage which makes the potential difference between two layers of the bilayer. Furthermore, it leads to the charge imbalance between the layers. In such approach, the Coulomb interaction between the layers is often neglected \cite{cite_17, cite_21}.
    The sufficiently large asymmetry gap, induced in the quasiparticle excitation spectrum, leads to the large charge-fluctuations in the system. Another treatment is based on the inclusion of the intralayer and interlayer Coulomb interactions. It has been shown recently \cite{cite_42, cite_43} that the proper inclusion of the interlayer Coulomb interaction in the Hamiltonian of the bilayer graphene, at half-filling in each layer, leads to a physically correct description of the excitonic effects in the system and describes well the variation of the Fermi level in the bilayer with respect to the interlayer interaction parameter. Meanwhile, it was shown that the intralayer Coulomb interaction induces a constant screening, in the BLG, by redefining the Fermi level in the system \cite{cite_42}.  
     
    We will calculate the real part of the ac conductivity $\sigma_{ij}(\Omega)$ in the BLG with the help of the retarded polarization function $\Pi^{R}_{ij}(\Omega)$ using the standard Kubo-Green-Matsubara formalism \cite{cite_50} 
    \begin{eqnarray}
    \Re{\sigma_{ij}(\Omega)}=\frac{\Im{\Pi^{R}_{ij}(\Omega)}}{\Omega},
    \label{Equation_4}
    \end{eqnarray}
    where 
    \begin{eqnarray}
    \Pi^{R}_{ij}(\Omega)=\Pi^{R}_{ij}(i\omega_m\rightarrow \Omega+i\eta^{+})
    \label{Equation_5}
    \end{eqnarray}
    is the real frequency retarded function, which is obtained after the analytical continuation of the bosonic Matsubara frequencies $i\omega_m$ into the upper half of the real axis in the complex plane, i.e., $i\omega_m\rightarrow \Omega+i\eta^{+}$ with $\eta^{+}$ being the infinitesimal positive constant, and the bosonic Matsubara frequencies $\omega_m$ are give as usual by $\omega_m=2\pi{m}/\beta$, where $m=0,\pm1,\pm2,...$, and $\beta=1/{k_{B}T}$. In turn, the current-current Matsubara correlation function $\Pi^{R}_{\rm ij}(i\omega_m)$ is given as \cite{cite_50} 
    \begin{eqnarray}
    \Pi^{R}_{\rm ij}(i\omega_m)=-\int^{\beta}_{0}d\tau e^{i\omega_m\tau}\left\langle T_{\tau}j_{i}(\tau)j_{j}(0)\right\rangle,
    \label{Equation_6}
    \end{eqnarray}
    where $\tau$ is the imaginary time, and $T_{\tau}$ is the chronological time ordering \cite{cite_50, cite_51}. The response function $\Pi^{R}_{\rm ij}(i\omega_m)$ corresponds to the particle-hole bubble, in terms of Matsubara
    Green's functions.
    The total current operator in the expression of the polarization function could be obtained from the expression in Eq.(\ref{Equation_2}), by functional differentiation of the tight-binding Hamiltonian with respect to $A_{i}(\mathbf{r})$ ($i$, being the index of the component  of the vector potential ${\bf{A}}$), i.e.
    \begin{eqnarray}
    j_{i}(\mathbf{r})=-\frac{\delta{H'}}{\delta{\left(A_{i}(\mathbf{r})/c\right)}}.
    \label{Equation_7}
    \end{eqnarray}
   Let's mention here, that the paramagnetic current operator $j_{i}(\mathbf{r})$ describes the in-plane electric current density, driven in the BLG layers by the external electric field. For the site-dependent paramagnetic current density operator, we get 
           \begin{eqnarray}
          j_{i}({\mathbf{r}})=-\frac{ie\gamma_0}{\hbar}\sum_{\bm{\mathit{\delta}}}\left[a^{\dag}_{\sigma}(\mathbf{r}+\delta)b_{\sigma}(\mathbf{r})\bm{\mathit{\delta}}_{i}-h.c.\right]
          \nonumber\\
          +\frac{ie\gamma_0}{\hbar}\sum_{\bm{\mathit{\delta}}'}\left[\tilde{a}^{\dag}_{\sigma}(\mathbf{r}+\delta)\tilde{b}_{\sigma}(\mathbf{r})\bm{\mathit{\delta}}'_{i}-h.c.\right],
           \label{Equation_8}
           \end{eqnarray}
       where the vectors $\bm{\mathit{\delta}}$ and $\bm{\mathit{\delta}}'$ are the nearest neighbor vectors in different layers of the bilayer graphene. The constituents of $\bm{\mathit{\delta}}$, for the bottom layer-1, are given by 
      $\bm{\mathit{\delta}}_{1}=\left({a_{0}}/{2\sqrt{3}},a_{0}/2\right)$, $\bm{\mathit{\delta}}_{2}=\left({a_{0}}/{2\sqrt{3}},-a_{0}/2\right)$, $\bm{\mathit{\delta}}_{3}=\left(-a_{0}/\sqrt{3},0\right)$, where $a_{0}=\sqrt{3}d$ is the sublattice constant (with $d$, being the carbon-carbon length in the graphene sheets). For the layer-2, we have obviously $\bm{\mathit{\delta}}'_{1}=\left(a_{0}/\sqrt{3},0\right)$, $\bm{\mathit{\delta}}'_{2}=\left(-{a_{0}}/{2\sqrt{3}},-a_{0}/2\right)$, $\bm{\mathit{\delta}}'_{3}=\left(-{a_{0}}/{2\sqrt{3}},a_{0}/2\right)$. We see that the vectors in the top layer-2 are oriented differently and it is not difficult to realise that $\bm{\mathit{\delta}}'=-\bm{\mathit{\delta}}$.
      Next, the current operator $j_{i}(\tau)$ in Eq.(\ref{Equation_6}) could be obtained after summing over the lattice sites in the operator given in Eq.(\ref{Equation_8}).  After transforming into the Fourier $\mathbf{k}$-space the expression in Eq.(\ref{Equation_8}) and after summing over the lattice sites positions in we get the following expression for the total current density operator
\begin{eqnarray}
 j_{i}(\tau)=\frac{1}{N}\sum_{\mathbf{k}}\left[{v}^{\ast}_{\mathbf{k}i}a^{\dag}_{1\mathbf{k}}(\tau)b_{1{k}}(\tau)+{v}_{\mathbf{k}i}b^{\dag}_{1\mathbf{k}}(\tau)a_{1\mathbf{k}}(\tau)\right.
 \nonumber\\
 \left.+{v}^{\ast}_{\mathbf{k}i}a^{\dag}_{2\mathbf{k}}(\tau)b_{2\mathbf{k}}(\tau)+{v}_{\mathbf{k}i}b^{\dag}_{2\mathbf{k}}(\tau)a_{2\mathbf{k}}(\tau)\right],
   \label{Equation_9}
   \end{eqnarray}
where

\begin{eqnarray}
 {v}_{\mathbf{k}i}=-\frac{i\gamma_0}{\hbar}\sum_{\bm{\mathit{\delta}}}\bm{\mathit{\delta}}_{i}e^{i{\mathbf{k}}\bm{\mathit{\delta}}}
 \label{Equation_10}
   \end{eqnarray}
is the electron velocity operator for the individual graphene sheet. 

\begin{figure}
 	\begin{center}
 		\includegraphics[scale=0.65]{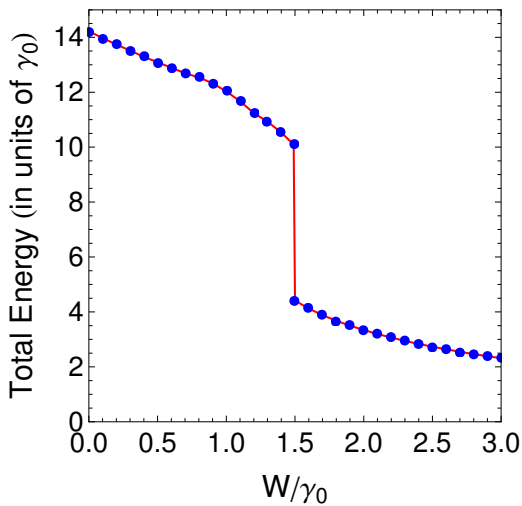}
 		\caption{\label{fig:Fig_1}(Color online) The total energy of the bilayer graphene structure, given via the interaction renormalized energy parameters in Eq.(\ref{Equation_21}). Zero temperature case is considered in the picture.}.
 	\end{center}
 \end{figure}

 It is worth to mention here that we perform the full 4-band calculation scheme, beyond the Dirac-cone approximation. We calculate the expression of the polarization function in Eq.(\ref{Equation_6}) by employing the Matsubara Green's function approach and we use the Wick averaging procedure for the fermionic field. 
  For the real part of the optical conductivity function, we obtain 
 \begin{eqnarray}
 &&\Re\sigma_{\rm xx}(\Omega)=\frac{\Im \Pi_{\rm xx}(\Omega)}{\Omega}=
 \nonumber\\
 &&{4\pi e^{2}v^{2}_{F}}\sum_{i,j=1}^{4}\int dx \rho_(x) P_{\rm ij}(x)\delta\left[\Omega+\varepsilon_{j}(x)-\varepsilon_{i}(x)\right]\times
 \nonumber\\
 &&\times\left[n_{F}(\mu+\Omega-\varepsilon_{i}(x))-n_{F}(\mu-\varepsilon_{i}(x))\right].
 \nonumber\\
 \label{Equation_11}
 \end{eqnarray}
 Here, the continuous variable $x$ of integration appears after the introduction of the non-interacting density of states $\rho(x)$ by 
 \begin{eqnarray}
 \rho(x)=\sum_{k}\delta(x-\gamma_{{\bf{k}}}),
  \label{Equation_12}
 \end{eqnarray}
 where $\gamma_{{\bf{k}}}$ is the tight-binding dispersion 
  \begin{eqnarray}
 \gamma_{{\bf{k}}}=\sum_{\bm{\mathit{\delta}}}e^{-i{\bf{k}}\bm{\mathit{\delta}}}.
 \label{Equation_13}
 \end{eqnarray}
 $v_{F}$ is the Fermi velocity, which relates to the intralayer hopping parameter $\gamma_0$, i.e., $v_{F}=\sqrt{3}a_0\gamma_0/2\hbar$.
 The function $P_{\rm ij}(x)$ in Eq.(\ref{Equation_11}) is the index-permutation function, given as    
 \begin{eqnarray}
 P_{\rm ij}(x)=\alpha_{i}(x)\beta_{j}(x)+\alpha_{j}(x)\beta_{i}(x),
 \label{Equation_14}
 \end{eqnarray} 
 where the dimensionless coefficients $\alpha_{i}(x)$ and $\beta_{i}(x)$ are defined as 
 \begin{eqnarray}
 \footnotesize
 \arraycolsep=0pt
 \medmuskip = 0mu
 \alpha_{i}(x)
 =(-1)^{i+1}\times
 \nonumber\\
 \times\left\{
 \begin{array}{cc}
 \displaystyle  & \frac{{\cal{P}}^{(3)}(\varepsilon_{i}(x))}{\left(\varepsilon_{1}(x)-\varepsilon_{2}(x)\right)}\prod^{}_{j=3,4}\frac{1}{\left(\varepsilon_{i}(x)-\varepsilon_{j}(x)\right)},  \ \ \  $if$ \ \ \ i=1,2,
 \newline\\
 \newline\\
 \displaystyle  & \frac{{\cal{P}}^{(3)}(\varepsilon_{i}(x))}{\left(\varepsilon_{3}(x)-\varepsilon_{4}(x)\right)}\prod^{}_{j=1,2}\frac{1}{\left(\varepsilon_{i}(x)-\varepsilon_{j}(x)\right)},  \ \ \  $if$ \ \ \ i=3,4,
 \end{array}\right.
 \nonumber\\
 \label{Equation_15}
 \end{eqnarray}
 where ${\cal{P}}^{(3)}(\varepsilon_{i}(x))$ is the polynomial of third order in $\varepsilon_{i}(x)$. Namely, we have
 \begin{eqnarray}	{\cal{P}}^{(3)}(\varepsilon_{i}(x))=\varepsilon^{3}_{i}(x)+\omega_{1}\varepsilon^{2}_{i}(x)+\omega_{2}(x)\varepsilon_{i}(x)+\omega_{3}(x)
 \label{Equation_16}
 \end{eqnarray}
 with the coefficients $\omega_{i}$, $i=1,...3$, given by 
 \begin{eqnarray}
 &&\omega_{1}=-2\mu^{\rm eff}_{2}-\mu^{\rm eff}_{1},
 \nonumber\\
 &&\omega_{2}(x)=\mu^{\rm eff}_{2}\left(\mu^{\rm eff}_{2}+2\mu^{ \rm eff}_{1}\right)-\left(\Delta+\gamma_1\right)^{2}-|{\gamma}_{0}|^{2}x^{2}, 
 \nonumber\\
 &&\omega_{3}(x)=-\mu^{\rm eff}_{1}\left(\mu^{\rm eff}_{2}\right)^{2}+\mu^{\rm eff}_{1}\left(\Delta+\gamma_1\right)^{2}+\mu^{ \rm eff}_{2}|{\gamma}_{0}|^{2}x^{2}.
 \nonumber\\
 \label{Equation_17}
 \end{eqnarray}
 The effective chemical potentials $\mu^{\rm eff}_{1}$ and $\mu^{\rm eff}_{2}$ that appear in Eq.(\ref{Equation_17}) are defined with the help of the interaction parameters $U$ and $V$, namely, we have $\mu^{\rm eff}_{1}=\mu+U/4$ and $\mu^{ \rm eff}_{2}=\mu+U/4+V$. Furthermore, they enters also in the expression of the effective Fermi level in the bilayer graphene, i.e., $\bar{\mu}=(\mu^{\rm eff}_{1}+\mu^{\rm eff}_{2})/2$. Next, for the coefficients $\beta_{i}(x)$, we have
 \begin{eqnarray}
 \footnotesize
 \arraycolsep=0pt
 \medmuskip = 0mu
 \beta_{i}(x)
 =(-1)^{i+1}\times
 \nonumber\\
\times\left\{
 \begin{array}{cc}
 \displaystyle  & \frac{{\cal{P}'}^{(3)}(\varepsilon_{i}(x))}{\left(\varepsilon_{1}(x)-\varepsilon_{2}(x)\right)}\prod^{}_{j=3,4}\frac{1}{\left(\varepsilon_{i}(x)-\varepsilon_{j}(x)\right)},  \ \ \  $if$ \ \ \ i=1,2,
 \newline\\
 \newline\\
 \displaystyle  & \frac{{\cal{P}'}^{(3)}(\kappa_{i}(x))}{\left(\varepsilon_{3}(x)-\varepsilon_{4}(x)\right)}\prod^{}_{j=1,2}\frac{1}{\left(\varepsilon_{i}(x)-\varepsilon_{j}(x)\right)},  \ \ \  $if$ \ \ \ i=3,4,
 \end{array}\right.
 \nonumber\\
 \label{Equation_18}
 \end{eqnarray}
 where ${\cal{P}'}^{(3)}(\varepsilon_{i}(x))$ is again a polynomial of third order in $\varepsilon_{i}(x)$, namely we have
 \begin{eqnarray}	
 {\cal{P}'}^{(3)}(\varepsilon_{i}(x))=\varepsilon^{3}_{i}(x)+\omega'_{1}(x)\varepsilon^{2}_{i}(x)+\omega'_{2}(x)\varepsilon_{i}(x)+\omega'_{3}(x)
 \nonumber\\
 \label{Equation_19}
 \end{eqnarray}
 with the coefficients $\omega'_{i}$, $i=1,...3$, given as 
 \begin{eqnarray}
 &&\omega'_{1}=-2\mu^{\rm eff}_{1}-\mu^{\rm eff}_{2},
 \nonumber\\
 &&\omega'_{2}(x)=\mu^{\rm eff}_{1}\left(\mu^{\rm eff}_{1}+2\mu^{\rm eff}_{2}\right)-|\tilde{\gamma}_{0}|^{2}x^{2},
 \nonumber\\
 &&\omega'_{3}(x)=-\mu^{\rm eff}_{2}\left(\mu^{\rm eff}_{1}\right)^{2}+\mu^{\rm eff}_{1}|\tilde{\gamma}_{0}|^{2}x^{2}.
 \nonumber\\
 \label{Equation_20}
 \end{eqnarray}
 We see that $\omega'_{1}$, $\omega'_{2}(x)$ and $\omega'_{3}(x)$ could be obtained from the coefficients $\omega_{1}$, $\omega_{2}(x)$ and $\omega_{3}(x)$ in Eqs.(\ref{Equation_17}), just by interchanging the effective chemical potentials $\mu^{(1)}_{\rm eff}\rightleftharpoons \mu^{(2)}_{\rm eff}$ and by setting simultaneously $\Delta+\gamma_1=0$. 
 The function $n_{F}(x)$ in the expression of the optical conductivity function is the usual Fermi-Dirac distribution function $n_{F}(x)=1/(e^{\beta{x}}+1)$. The principal attention in Eq.(\ref{Equation_11}) has to be put on the energy parameter $\varepsilon_{i}(x)$ with $i=1,...4$. Those parameters are the quasiparticle energy dispersions in the BLG, which are renormalized by the Coulomb interaction effects in the BLG, and form the four band structure of the BLG system. They are given as (for more details on the subject see in Ref.\onlinecite{cite_42})
 \begin{align}
 	\varepsilon_{1,2}(x)=-\frac{1}{2}\left[\Delta+\gamma_{1}\pm\sqrt{\left(W-\Delta-\gamma_{1}\right)^{2}+4|\tilde{\gamma}_{0}|^{2}}x^{2}\right]
 	\nonumber\\
 	+\bar{\mu},
 	\nonumber\\
 	\varepsilon_{3,4}(x)=-\frac{1}{2}\left[-\Delta-\gamma_{1}
 	\pm\sqrt{\left(W+\Delta+\gamma_{1}\right)^{2}+4|\tilde{\gamma}_{0}|^{2}x^{2}}\right]
 	\nonumber\\
 	+\bar{\mu}.
 	\nonumber\\
 	\label{Equation_21}
 \end{align}
 Indeed, the parameters $\varepsilon_{2}(x)$ and $\varepsilon_{3}(x)$ are the low energy conduction and valence bands, and $\varepsilon_{1}(x)$ and $\varepsilon_{4}(x)$ are the split valence and conduction bands in the usual tight-binding notations \cite{cite_37, cite_42}. 
 The excitonic gap parameter $\Delta$ in Eq.(\ref{Equation_21}) is defined as 
 \begin{eqnarray}
 \Delta=W\left\langle {b}^{\dag}({\bf{r}}\tau)\tilde{a}({\bf{r}}\tau)\right\rangle.
 \label{Equation_22}
 \end{eqnarray}
 In  Fig.~\ref{fig:Fig_1}, we have presented the total energy of the interacting BLG system as a function of the interlayer normalized Coulomb interaction parameter $W$. The zero temperature case is considered in the picture. We see that at the critical value of the interlayer coupling parameter $W_{c}=1.49\gamma_0=4.47$ eV, the total energy of the system has a large jump into the lower bound states. This corresponds well with the discussions, given in Refs.\onlinecite{cite_42, cite_43}, concerning the exact chemical potential and the Fermi level in the BLG. Especially, it has been shown in Ref.\onlinecite{cite_42} that the Fermi energy passes into its upper bound solution at $W_{c}=1.49\gamma_0$, by defining a new type of the charge neutrality point in the interacting bilayer graphene system. The energy jump, given in Fig.~\ref{fig:Fig_1} is also a demonstration of the existence of broken symmetry state, leading to the formation of the excitonic insulator and the excitonic condensate states in the BLG (when augmenting the interaction parameter), as it was well pointed out in Ref.\onlinecite{cite_43}.  
 %
 %
  Indeed, the intralayer Coulomb interaction parameter $U$ redefines the Fermi level in the BLG, and, for the non-zero values of the interlayer interaction parameter, the Fermi level in the bilayer graphene is given as $\varepsilon_{F}=\bar{\mu}=\mu+\kappa{U}+0.5V$ (with $\kappa=0.25$) \cite{cite_42, cite_43}, where the chemical potential $\mu$ could be calculated after the self-consistent equations given in Ref. \onlinecite{cite_42}. For a detailed analysis about the excitonic modification of the electronic band structure, and also the behavior of the chemical potential, Fermi level, and the excitonic gap parameter, we refer the reader to the works in Refs.\onlinecite{cite_42, cite_43}. 
  Within the same context, it has been shown in Ref.\onlinecite{cite_43}, how the variation of the Fermi level $\bar{\mu}$ with respect to the interlayer Coulomb interaction parameter $W$ drives the BLG system into the semiconducting (or insulating) state from the semimetallic limit one, leading also to the enhancement of the Bardeen-Cooper-Schrieffer (BCS)-Bose-Einstein-Condensate (BEC) excitonic crossover in the BLG. A threshold mechanism of creation of the band gap $E_g$ and the hybridization gap $\Delta_{H}$ in the single-particle excitation spectrum has been also discussed there, in details. It is particularly important to notice here that the opening of the excitonic band gap, discussed in the mentioned papers, is totally different from the formation of the asymmetry band gap between the low energy conduction and valence bands (known like the “sombrero” shape, \cite{cite_1, cite_2}) in the presence of the external electric field asymmetry between the layers of the BLG \cite{cite_28}. In contrast to the usual splitting bands behavior \cite{cite_28}, here we have no intraband splitting at the Dirac's point $K$, between the low-energy valence and split bands. This is true also for the very high values of the interlayer interaction parameter $W$. 
  The imaginary part of the optical conductivity function $\sigma_{\rm xx}(\Omega)$ could be easily calculated using the Kramers-Kronig relation, which relates the real and imaginary parts of the complex conductivity function. It is given as 
 \begin{eqnarray}
 {\Im\sigma_{\rm xx}}(\Omega)=-\frac{2\Omega}{\pi}\int^{\infty}_{0}d\Omega'\frac{\Re\sigma_{\rm xx}(\Omega')}{\Omega'^{2}-\Omega^{2}},
 \label{Equation_23}
 \end{eqnarray}  
 where a special attention have to be paid to the singularity points $\Omega'=\pm\Omega$ when performing the numerical integration in Eq.(\ref{Equation_23}).

\section{\label{sec:Section_4} The optical properties of the BLG}
%

 We consider here the suspended bilayers, and the interaction effects have a spectacular impact on the conductivity spectrum because the dielectric environment portion of the screening is absent.
%
\begin{figure}
	\begin{center}
		\includegraphics[scale=0.65]{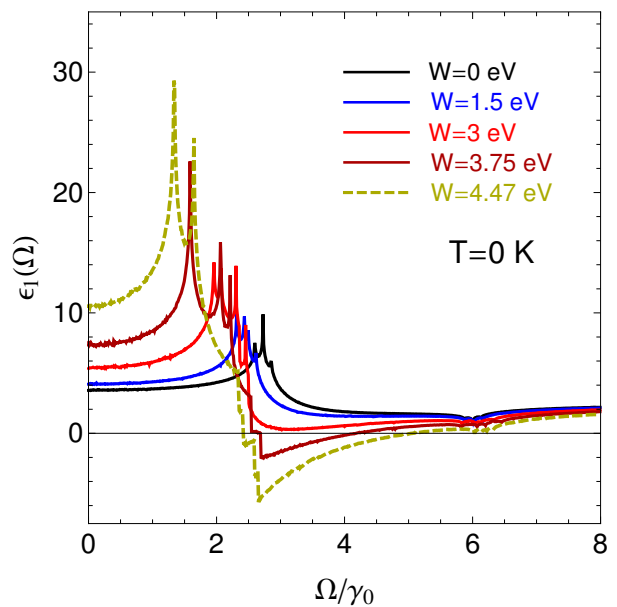}
		\caption{\label{fig:Fig_2}(Color online) The real part of the dielectric function, given in Eq.(\ref{Equation_24}), for the case $T=0$ K. The interlayer hopping parameter is set at $\gamma_1=0.384$ eV, and different values of the interlayer interaction parameter (from zero up to $W_c$) are considered.}.
	\end{center}
\end{figure} 
%
\begin{figure}
	\begin{center}
		\includegraphics[scale=0.65]{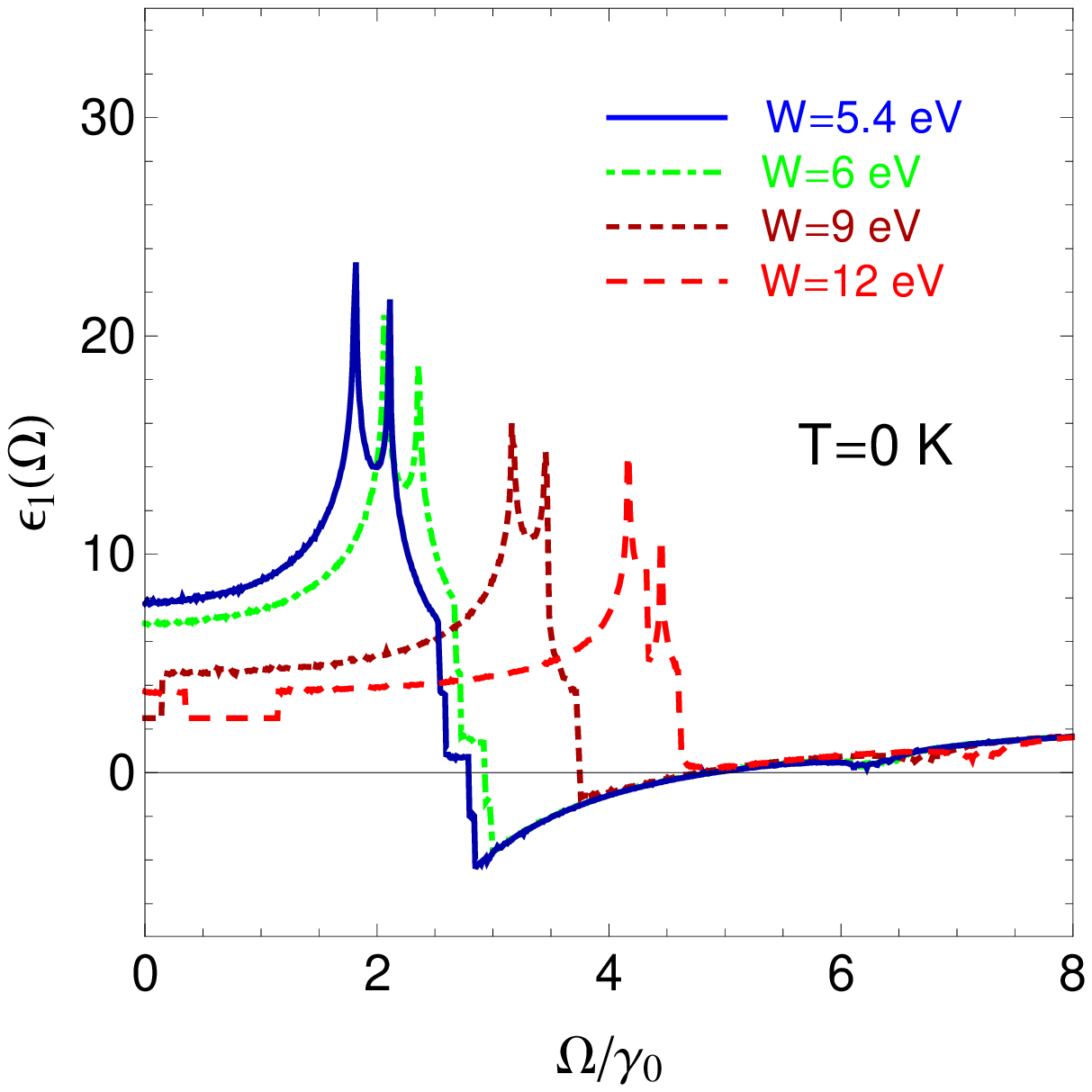}
		\caption{\label{fig:Fig_3}(Color online) The real part of the dielectric function, given in Eq.(\ref{Equation_24}), for the case $T=0$ K. The interlayer hopping parameter is set at $\gamma_1=0.384$ eV, and different values of the interlayer interaction parameter (from intermediate up to very high) are considered.}.
	\end{center}
\end{figure} 
%
\begin{figure}
	\begin{center}
		\includegraphics[scale=0.65]{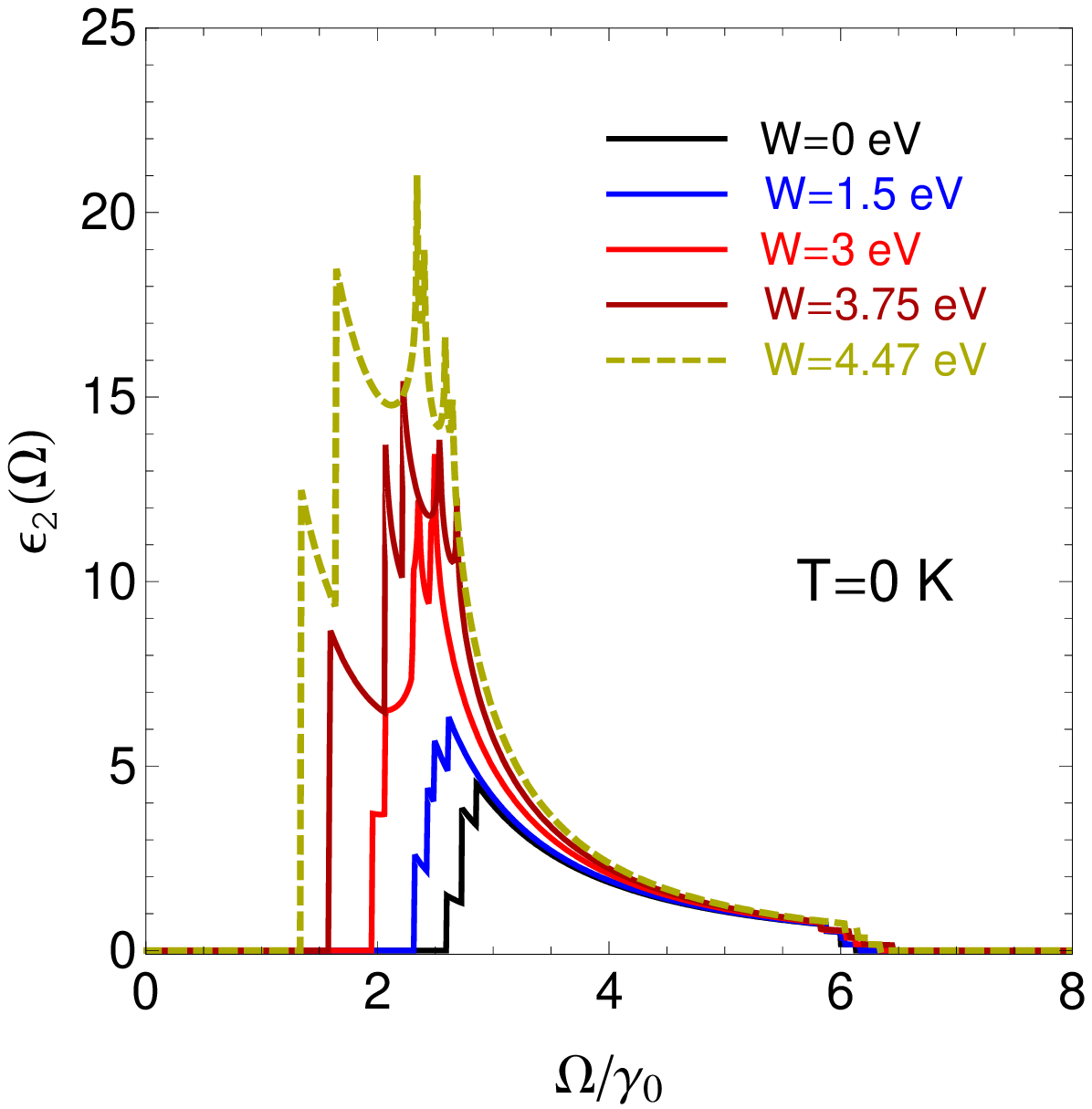}
		\caption{\label{fig:Fig_4}(Color online) The imaginary part of the dielectric function, given in Eq.(\ref{Equation_24}), for the case $T=0$ K. The interlayer hopping parameter is set at $\gamma_1=0.384$, eV and different values of the interlayer interaction parameter (from zero up to $W_c$) are considered.}.
	\end{center}
\end{figure} 
%
\begin{figure}
	\begin{center}
		\includegraphics[scale=0.65]{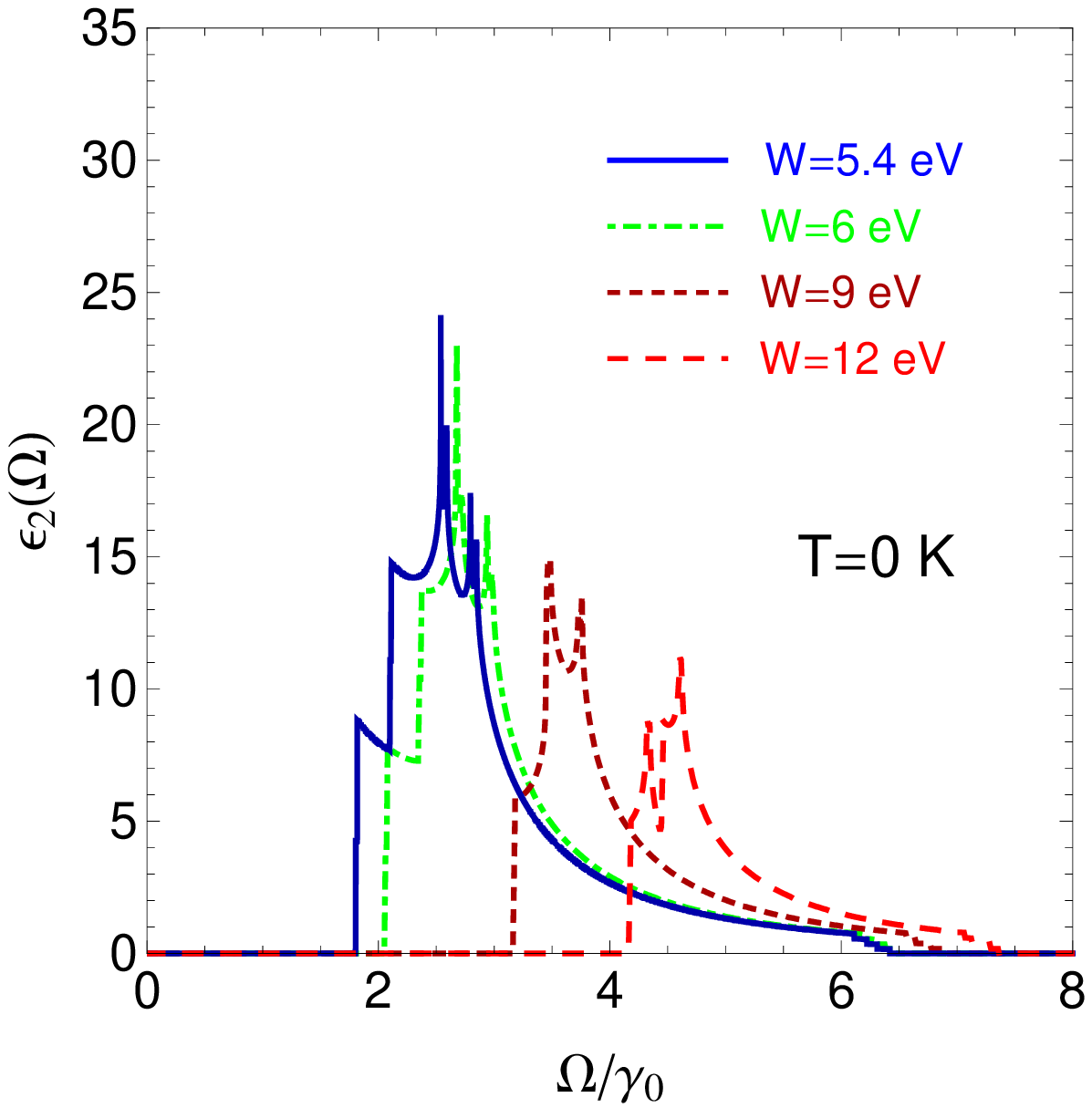}
		\caption{\label{fig:Fig_5}(Color online) The imaginary part of the dielectric function, given in Eq.(\ref{Equation_24}), for the case $T=0$ K. The interlayer hopping parameter is set at $\gamma_1=0.384$ eV, and different values of the interlayer interaction parameter (from intermediate up to very high) are considered.}.
	\end{center}
\end{figure} 
%
\begin{figure}
	\begin{center}
		\includegraphics[scale=0.65]{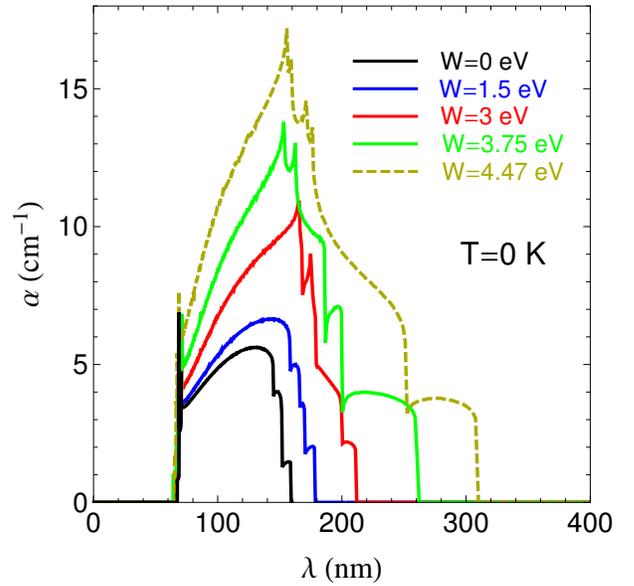}
		\caption{\label{fig:Fig_6}(Color online) The absorption coefficient in the bilayer graphene, for the case $T=0$ K. The interlayer hopping parameter is set at $\gamma_1=0.384$ eV, and different values of the interlayer interaction parameter (from zero up to $W_c$) are considered.}.
	\end{center}
\end{figure} 
%
The optical properties of the suspended bilayer graphene are directly connected to the frequency dependent complex dielectric function $\epsilon_{\rm xx}(\Omega)=\epsilon_{1}(\Omega)+i\epsilon_{2}(\Omega)$. Particularly the imaginary part of it gives the information about the optical absorption spectra in the BLG. For the applied in-plane electric field in the $x$-direction, the dielectric function $\epsilon_{\rm xx}(\Omega)$ is directly related to the complex optical conductivity function $\sigma_{\rm xx}(\Omega)=\Re(\sigma_{\rm xx})+i\Im(\sigma_{\rm xx})$ by the following relation \cite{cite_52} 
\begin{eqnarray}
{\epsilon}_{\rm xx}(\Omega)=\epsilon_{\rm g}+4\pi{i}\frac{\sigma_{\rm xx}(\Omega)}{\Omega{d_n}},
\label{Equation_24}
\end{eqnarray}  
where $\epsilon_{\rm g}=2.5$ and $d_n=2.36$ is the normalized interlayer distance (the interlayer distance is normalized to the interatomic separation in the given graphene sheet of the BLG, i.e., we take $d_n=d/a$, where $d=0.335$ nm is the interlayer separation in the BLG and $a_0=0.142\sqrt{3}$ nm is the sublattice constant in the graphene sheet). The imaginary part of ${\varepsilon}_{\rm xx}(\Omega)$ leads to absorption loss if it is positive and absorption gain if it is negative.  We see in Fig. ~\ref{fig:Fig_2} that the peaks positions in the real part of the dielectric function get red-shifted for the interlayer interaction interval $W\in[0;W_c]$, while the blue-shifted curves in Fig.~\ref{fig:Fig_3} corresponds to the strong interlayer interaction regime $W>W_c$. The absorption loss spectra in the BLG, for the whole interaction range, is presented in Figs.~\ref{fig:Fig_4} and ~\ref{fig:Fig_5}. For the strong coupling regime, the amplitudes of the optical loss spectra are decreasing with the increase of $W$, suggesting the enhancement of the strong optical transmission in the BLG in the very narrow range of the photon's wavelengths. 

\begin{figure}
	\begin{center}
		\includegraphics[scale=0.65]{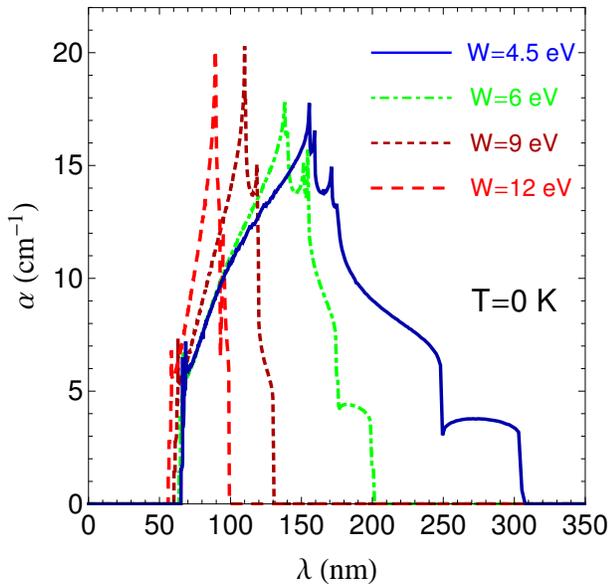}
		\caption{\label{fig:Fig_7}(Color online) The absorption coefficient in the bilayer graphene, for the case $T=0$ K. The interlayer hopping parameter is set at $\gamma_1=0.384$ eV, and different values of the interlayer interaction parameter (from intermediate up to very high) are considered.}.
	\end{center}
\end{figure}

\begin{figure}
	\begin{center}
		\includegraphics[scale=0.65]{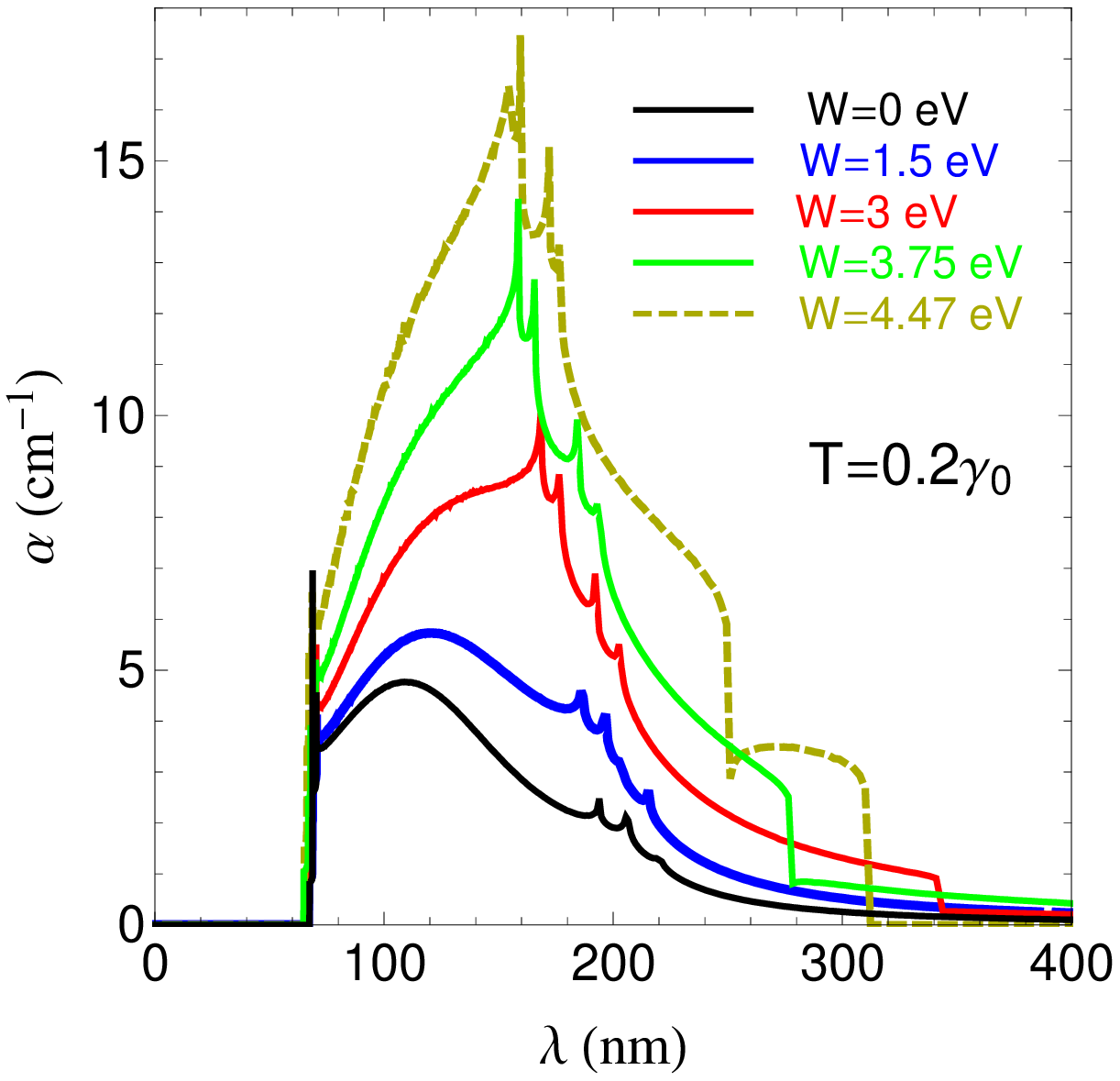}
		\caption{\label{fig:Fig_8}(Color online) The absorption coefficient in the bilayer graphene, for the case $T=0.2\gamma_0$. The interlayer hopping parameter is set at $\gamma_1=0.384$ eV, and different values of the interlayer interaction parameter (from zero up to $W_c$) are considered.}.
	\end{center}
\end{figure}

We have calculated also the absorption coefficient $\alpha(\Omega)$ and the real part of the complex optical refractive index $n(\Omega)$ for all mentioned values of the interlayer interaction strength, by fixing the intralayer Coulomb interaction at the value $U=2\gamma_0=6$ eV. For this, we used the well-known relations \cite{cite_53}
\begin{eqnarray}
&&\alpha(\Omega)=\frac{\Omega\varepsilon_{2}(\Omega)}{n(\Omega)},
\nonumber\\
&&n(\Omega)=\sqrt{\frac{\sqrt{\varepsilon^{2}_{1}(\Omega)+\varepsilon^{2}_{2}(\Omega)}+\varepsilon_{1}(\Omega)}{2}}.
\label{Equation_25}
\end{eqnarray} 
We see in Fig.~\ref{fig:Fig_6} that the maximal absorption at $W=0$, corresponds to the incident wavelength of order $\lambda=131.558$ nm, situating in the UV-C region of the UV spectrum. It corresponds to the high photon energy of the order of $\sim9.42$ eV. The optical absorption threshold corresponding to the first peak in the optical absorption spectrum in Fig.~\ref{fig:Fig_6} is at the photon's wavelength of order $\lambda=158.583$ nm (with the energy of the order of $7.81$ eV), which is again in the UV-C range of the photon's spectrum. When augmenting the interlayer interaction parameter $W$, the absorption coefficient increases drastically and gets red-shifted (when $W\in[0;W_c]$) at the positions of the highest absorption attenuations. It is interesting to notice that the wavelengths values corresponding to the absorption coefficient in Fig.~\ref{fig:Fig_6} cover the narrow range of the spectrum from extreme UV-C up to near UV region, and also the absorption threshold is unchanged in that case. Contrary, for the strong interlayer interaction parameter the absorption peaks positions get blue-shifted attaining the region of the extreme UV lengths (see the red curve, in the left side of the picture in Fig.~\ref{fig:Fig_7}). In Fig.~\ref{fig:Fig_8}, we give the absorption coefficient for the very high value of temperature $T=0.2\gamma_0$ and we observe only the small changes in the spectral curves. Namely, the amplitudes of the absorption lines are slightly diminished in this case and the entire spectrum is also spread over the higher values of the incident photon wavelengths. This remains true for both, intermediate and strong values of the interlayer interaction strengths.         
 
\begin{figure}
	\begin{center}
		\includegraphics[scale=0.65]{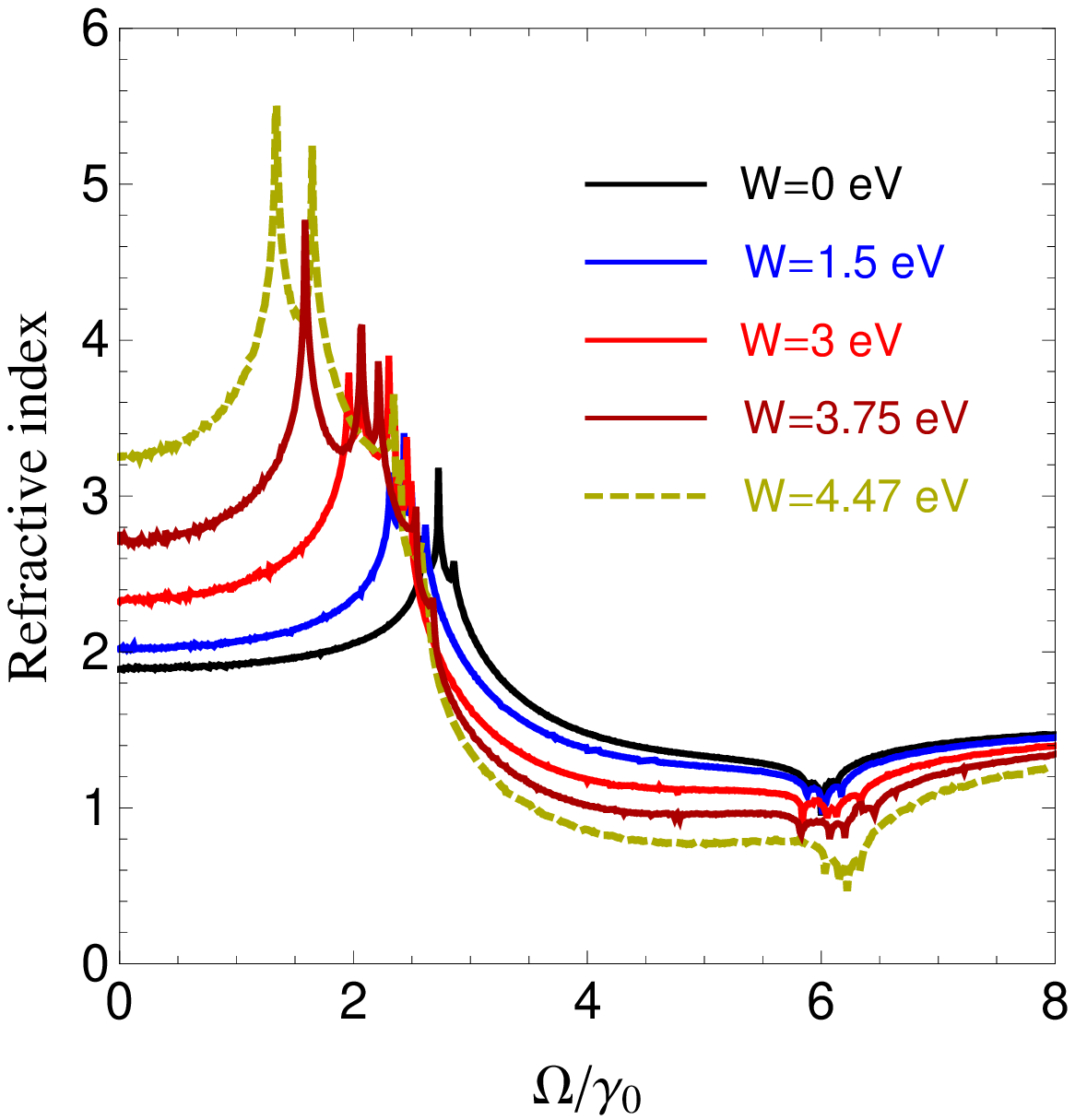}
		\caption{\label{fig:Fig_9}(Color online) The refractive index in the bilayer graphene, for the case $T=0$ K. The interlayer hopping parameter is set at $\gamma_1=0.384$ eV, and different values of the interlayer interaction parameter (from zero up to $W_c$) are considered.}.
	\end{center}
\end{figure}

\begin{figure}
	\begin{center}
		\includegraphics[scale=0.65]{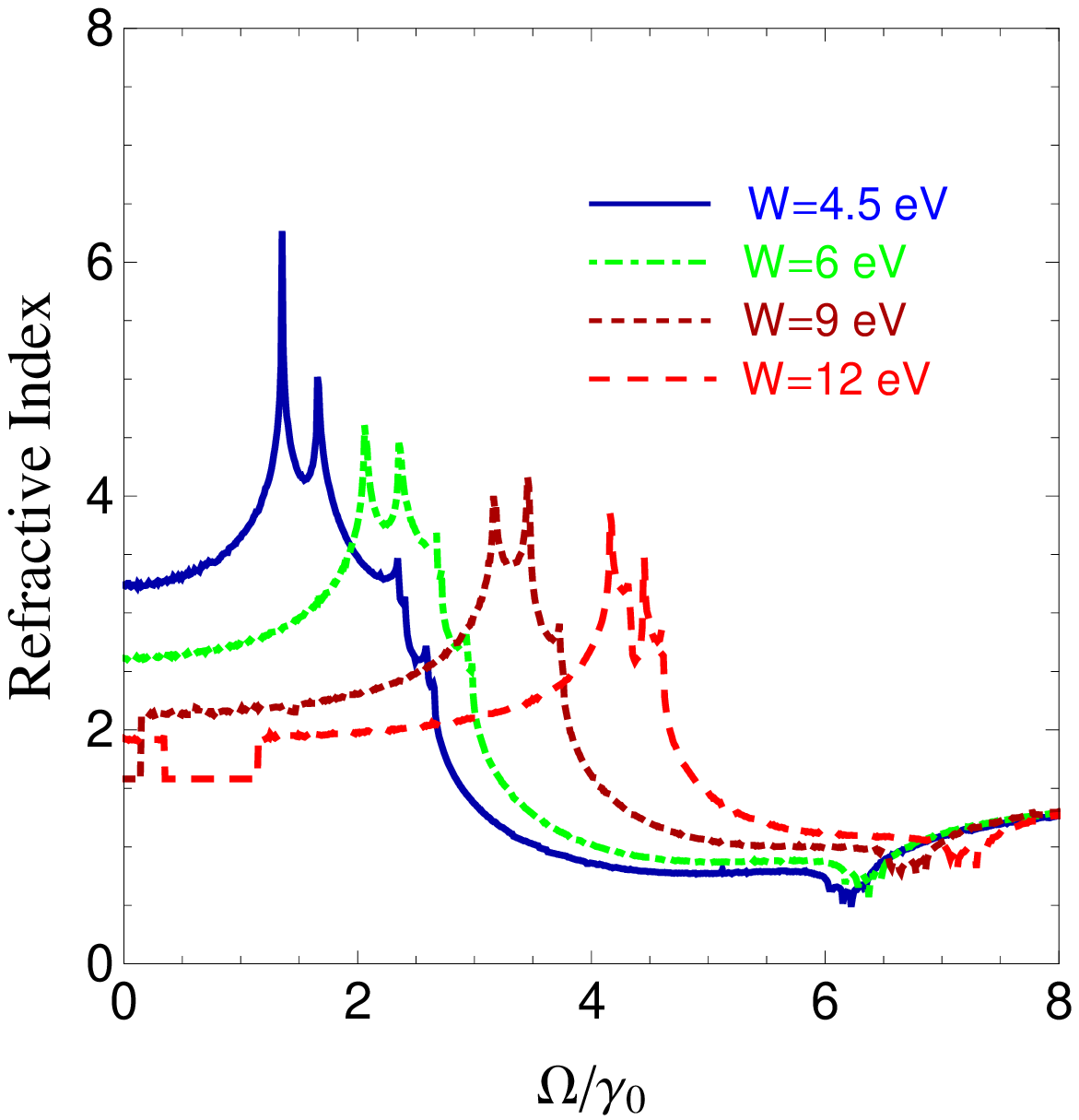}
		\caption{\label{fig:Fig_10}(Color online) The refractive index in the bilayer graphene, for the case $T=0$ K. The interlayer hopping parameter is set at $\gamma_1=0.384$ eV, and different values of the interlayer interaction parameter (from intermediate up to very high) are considered.}.
	\end{center}
\end{figure}

The refractive index of photonic materials such, as two-dimensional materials, varies with the
photon energy and/or wavelength of the applied electric field. After getting the real and imaginary parts of the dielectric function in the BLG, we have calculated the real part $n(\Omega)$ of the complex refractive index parameter in the bilayer graphene. In Figs.~\ref{fig:Fig_9} and ~\ref{fig:Fig_10}, we give the real part of refractive index of the bilayer graphene, for different values of the Coulomb interaction parameter. It has complicated behavior as a function of the normalized incident photon energies $\Omega/\gamma_0$. We have shown that the refractive increases continuously for $0<W\leq W_c$ in the frequency region $0<\Omega<2.59\gamma_0$ ($0<\Omega<7.78$ eV). For higher frequencies, it decreases with increasing of $W$. When passing to the strong interlayer coupling limit $W>W_c$ (see in Fig.~\ref{fig:Fig_10}), the inverse effect takes place due to the strong blue-shift effect, i.e., the refractive index decreases with increasing of $W$ in the domain $0<\Omega<7.78$ eV and increases nearly continuously with increasing $W$ when $\Omega>7.78$ eV. We observe in Fig.~\ref{fig:Fig_9} that the values of refractive index, corresponding to $W=0$ (see the black curve), is of order of $n=1.9$, and $n=2.03$ for $W=0.5\gamma_0=1.5$ eV (see the blue curve in the picture). Those values remain practically unchanged for a very large interval of the incident photon wavelengths $\lambda\in[275.5; 826.6]$ nm, covering the UV-B and the visible part of the spectrum. This is not true for the case of the intermediate values of the Coulomb interaction parameter $W\in[\gamma_0; W_c]$ and the refractive index varies spectacularly when changing the parameter $W$. Contrary, for the very large interaction strengths (see the curves, corresponding to the cases $W=3\gamma_0=9$ eV and $W=4\gamma_0=12$ eV), the refractive index takes again the constant values, for a very large region of $\lambda$ and further modifications occur only at the very high-frequency region, corresponding to the far UV range of the spectrum.
It is remarkable to note here that for the non-interacting BLG, i.e., when $W=0$, the value $n(\Omega)=1.9$ found in the very low-$\Omega$ region coincides well with the value of the refractive index at the large interlayer separation distances, given in Ref.\onlinecite{cite_14}, where the optical properties of AA-stacked bilayer graphene have been studied using the density functional theory and when considering the case of the in-plane electric field polarization. There is also a good agreement with Ref.\onlinecite{cite_14} at the very high frequencies, for which $n(\Omega)\sim 1$. The further effects of deviations from our results, in the case of the finite and large frequency regions, are probably due to the stacking order and the effect of interactions, which have been neglected completely in Ref.\onlinecite{cite_14}. A good agreement with the results in Ref.\onlinecite{cite_14} is achieved also concerning the plasmon excitation spectra, especially at the high electron loss energies, as we will see later on in the paper. We see in Fig.~\ref{fig:Fig_9} that when augmenting the interlayer interaction parameter the refractive index increases and also the excitonic peaks in the spectrum are displacing into the region of small normalized frequencies. This red-shift effect of the peak positions turns, furthermore, into the blue-shift displacement of the refractive index peaks positions when $W>W_c$. At the value $\lambda=634$ nm of the incident wave, we have $\Omega/\gamma_0=0.652$ (or in energy units $\Omega=1.956$ eV) and the refractive index is $n=2.785$ for the case $W=1.25\gamma_0$. At this values of $W$ the excitonic gap parameter attains its largest possible value (see in Ref.\onlinecite{cite_42}. 
 
 Recently, the refractive index of graphene has been measured by two independent experimental techniques for determining both the real and imaginary refractive index values of graphene \cite{cite_54}. Additionally, the authors have calculated the real and imaginary parts of the complex dielectric function of pristine graphene by using the density functional theory. For the weak electric field oscillating parallel to the graphene layer they have found $\sqrt{\epsilon_{\rm xx}}=n+i\kappa=2.71+i1.41$ at the same incident wavelength of order of $\lambda=634$ nm, corresponding to $\Omega=1.96$ eV. Thus $n=2.71$, while the measured refractive index data for different samples shows that $n=2.65+i1.27$. The obtained theoretical value fits well with our result for $W=1.25\gamma_0=3.75$ eV.   
 
The obtained experimental value of the refractive index is also attainable from our theory if we consider the slightly higher value of the interlayer Coulomb interaction parameter. Particularly, for the value $W=2\gamma_0=6$ eV (which is already the strong coupling limit of the BLG, when the condensate states are developing \cite{cite_43}), we get $n=2.672$ at $\lambda=634$ nm, which is very close to the observed experimental value in Ref.\onlinecite{cite_54}. It is remarkable to note, that at the mentioned value of the incident wavelength, the obtained values of the refractive index remain practically the same for the very high temperature $T=0.2\gamma_0$ (the corresponding curves are not presented in the presented here). In this case, we obtain $n=2.93104$ for  $W=1.25\gamma_0$ and $n=2.73891$ for the case $W=2\gamma_0$. Slightly higher values of the refractive index, in this case, concern the low energy photon region of the spectrum, While in the intermediate range $\Omega\in[\gamma_0;3\gamma_0]$, the refractive index coefficient is considerably smaller than for the case $T=0$. Thus, we can conclude that the effect of temperature is negligibly small when determining the refractive index coefficient. The important effect of our calculations is that the refractive index in the BLG could be also very high ($n>3$), in two different regimes, namely, for the frequency range $\Omega\in[3;9]$ eV (see in Fig.~\ref{fig:Fig_9}), corresponding to the interlayer interaction interval $W\in[0;4.47]$ eV, and in the case of frequency range $\Omega\in [3;15]$ eV (see in Fig.~\ref{fig:Fig_10}), corresponding to $W\in [4.5;12]$ eV. Thus, by suggesting experimentally the refractive index values in the BLG, one should properly include the effects of the interlayer Coulomb interaction, in order to get the sufficient convergence with the theory. It is very important to mention here that the peaks in the spectrum of the refractive index, obtained in Figs.~\ref{fig:Fig_9} and ~\ref{fig:Fig_10}, show explicitly the presence of the strong excitonic effect in the half-filled BLG system. For the case $0<W<W_c$ eV the peaks positions are situated in the wavelength region $\lambda\in[137.7; 413.3]$ nm, thus covering practically the whole UV region of the spectrum. This is consistent with the discussion about the optical conductivity spectrum in the BLG. Meanwhile, for $W\geq4.5$ eV we have $\lambda\in[82.6; 413.3]$ nm, thus going deep further into the extreme UV side. The general consequence gained here is the fact that the excitonic effects in the bilayer graphene take place mainly in the UV region of the photon spectrum, corresponding to the relatively high incident photon's energies. 

The spectrum of the dielectric function in the BLG is important in order to determine the electron energy loss function, i.e., the plasmon excitation spectrum in the BLG 
\begin{eqnarray}
&&\Im(-\frac{1}{\epsilon(\Omega)})=\frac{\epsilon_{2}(\Omega)}{\epsilon^{2}_{1}(\Omega)+\epsilon^{2}_{2}(\Omega)}
\label{Equation_26}
\end{eqnarray} 
 Particularly,  measuring optical properties with EEL spectroscopy offers advantages of better spatial resolution and it extends to higher photon's energies. The maximum of the loss-function is identified as occurring at the point where $\epsilon_{1}(\Omega)$ crosses zero with
 a positive slope, i.e., when $d\epsilon_{1}/d\Omega>0$, and when $\epsilon_{2}(\Omega)<<1$ but with a negative slope, i.e., when $d\epsilon_{2}/d\Omega<0$. It is clear from pictures in Figs.~\ref{fig:Fig_2}, ~\ref{fig:Fig_3}, ~\ref{fig:Fig_4}, ~\ref{fig:Fig_5} that the plasmonic excitation peaks will be situated at the very high-frequency regions and the BLG will be transparent only at the very high frequencies of the electric field component of the light. For the low frequencies, the BLG will show strong screening effects. 
 
\begin{figure}
	\begin{center}
		\includegraphics[scale=0.65]{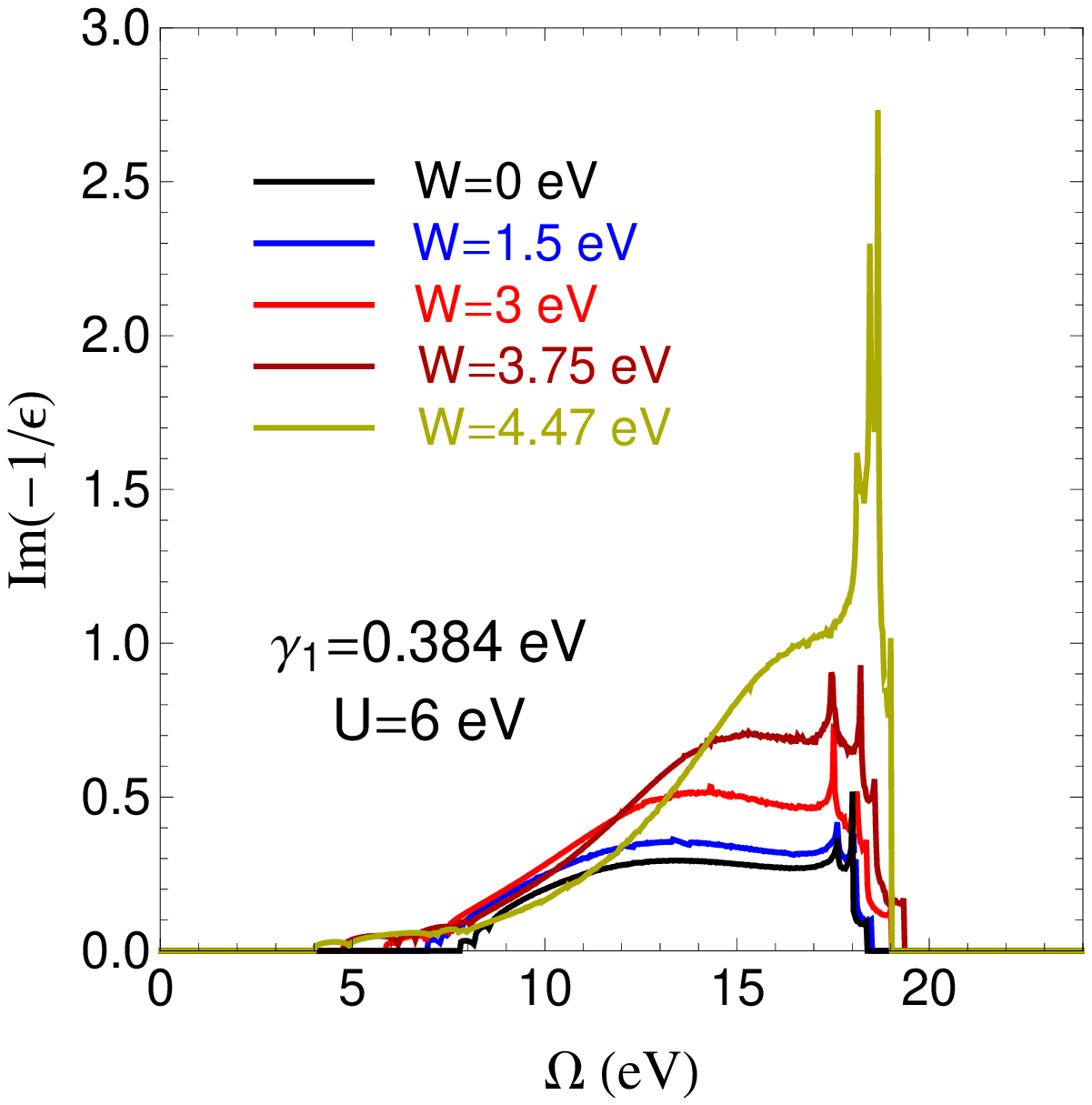}
		\caption{\label{fig:Fig_11}(Color online) Electron energy loss function for bilayer graphene at different interlayer interaction regimes. The zero temperature case is considered in the picture. }.
	\end{center}
\end{figure}

In Fig.~\ref{fig:Fig_11}, we have shown the electron energy loss (EEL) function calculated with Eq.(\ref{Equation_26}), for different values of the interaction parameter $W$. We see that the sharp plasmonic resonance peaks, for both interacting and noninteracting (W=0) BLG, are well situated in the very high-frequency interval $\Omega\in[16,20]$ eV. This result agrees well with the EEL function behavior at the small interlayer distances, obtained with the help of the ab-initio calculations and for the in-plane polarized electric field for the AA-stacked BLG (see in Ref.\onlinecite{cite_14}). Particularly, we should notice that we have a gradual reduction of the low-energy plasmonic resonance peaks (see the weakly pronounced peaks structures in the low-energy part o the spectrum in Fig.~\ref{fig:Fig_11}), which is probably due to the stacking order of BLG considered here and also the considerable interaction effects, which have been neglected completely in Ref.\onlinecite{cite_14} (particularly, this concerns also to the excitonic interaction effects in our AB-BLG structure). It is remarkable, that the low-energy plasmonic peaks in the EEL spectrum get also suppressed (see in Ref.\onlinecite{cite_14}) when decreasing the interlayer separation $d$, and thus by increasing the effects of interaction between the layers in the BLG. Let's mention also that the increase of the interlayer coupling leads to the gradual increase of the plasmonic excitations in the system in the high-frequency region: an effect which is completely absent in the results in Ref.\onlinecite{cite_14}. Meanwhile, it is important to observe that the interaction effects in the BLG lead also to a red-shift effect in the low-energy part of plasmonic excitations (see the visible red-shifted lines for the low-energy part of EEL, i.e., when $\Omega\in(4;8)$ eV).  
%
%
%
%
%
%
\section{\label{sec:Section_5} Concluding remarks}
%
We have calculated the optical properties of the interacting bilayer graphene. The full interaction bandwidth has been taken into account, and we have used the accurate four-band model of the BLG, which involves the contributions coming from all four bands, by avoiding various simplifications related to the low-energy effective considerations, given in the existing literature on the same subject.  
We observed that the optical properties of the bilayer graphene become more spectacular when including the interlayer interaction parameter $W$. We have calculated theoretically the optical conductivity as a necessary prerequisite, in order to consider the excitonic effects in the system. We have shown that all optical properties in the interacting bilayer graphene are strongly related to the behavior of the chemical potential and the exact Fermi level in the BLG. 

At the new charge neutrality point, governed by the excitonic pairing and condensation in the BLG, and corresponding to the interlayer coupling strength $W_c=1.49\gamma_0=4.47$ eV, the chemical potential passes into its upper bound solution. In this case, the optical absorption spectra get switched inversely from the red-shifted lines into the blue-shifted ones. This effect affects the whole optical spectra and the physical properties of the bilayer graphene. The dielectric response function, absorption coefficient and the refractive index of the BLG system get different and opposite behaviors when the Fermi energy passes from its lower bound solution to the upper bound one, after modifying the interlayer Coulomb interaction parameter. The optical absorption coefficient shows the largest absorption peaks, situated in the far UV range of the photon's wavelengths. By calculating the real part of the complex refractive index, we have shown that for the photon's wavelengths starting from UV-C up to visible region, the refractive index of the bilayer graphene increases continuously with increasing the interlayer coupling parameter, up to the charge neutrality value $W_c$, while, in the deep and shorter UV-C side of the incident wavelengths, the refractive index decreases when increasing $W$. For the values of $W$ above the charge neutrality value ($W>W_c$), the refractive index shows the opposite behavior, i.e. the refractive index decreases with $W$ for the mentioned interval of the photon's wavelengths, while it is increasing for the shorter wavelengths. We have calculated also the electron energy loss function, taking into account the interaction effects in our BLG system. Considering again various interlayer coupling regimes, we have obtained the high energy plasmonic peaks in the BLG. In the low-energy part of the spectrum, the amplitudes of the plasmonic peaks get gradually reduced due to the excitonic interaction effects and also the stacking ordering in our bilayer graphene system. At different stages of our calculations, we have compared our results with the exact ab-initio calculation and experimental results. We have explained in details the common features and deviations from those results. 

Presented work could be very interesting for correct estimations of various important optical parameters in the interacting bilayer graphene. Since the growing experimental interests in the optical properties of the graphene-based materials and heterostructures, the correct understanding of the interaction effects and their leading renormalizations on different physical quantities in the bilayer graphene is straightforward for correct understanding the behavior of the measured physical quantities. We hope that the theory, given here, will meet the most demanded criteria and situations concerning the physics and the measurements of different optical parameters in the bilayer graphene. 
\appendix
   	

%

\begin{thebibliography}{10}
%
\bibitem{cite_1} A. H. Castro Neto, F. Guinea, N. M. R. Peres, K. S. Novoselov, A. K. Geim, Rev. Mod. Phys. 81 109 (2009).
\bibitem{cite_2} E. V. Castro, et al. Phys. Rev. Lett 99 216802 (2007).
\bibitem{cite_3} J. Nilsson, A. H. Castro Neto, F. Guinea and N. M. R. Peres Phys. Rev. B 76 165416 (2007).
\bibitem{cite_4} K. Novoselov, A. K. Geim, S. V. Morozov, D. Jiang, M. I.
Katsnelson, I. V. Grigorieva, S. V. Dubonos, and A. A. Firsov, Nature London 438, 197 (2005).
\bibitem{cite_5} D. S. L. Abergel and Vladimir I. Fal'ko,  Phys. Rev. B 75, 155430 (2007).
\bibitem{cite_6} E. V. Gorbar, V. P. Gusynin, A. B. Kuzmenko, and S. G. Sharapov, Phys. Rev. B 86, 075414 (2012).
\bibitem{cite_7} J. Nilsson, A. H. Castro Neto, F. Guinea, and N. M. R. Peres, Phys. Rev. Lett. 97, 266801 (2006). 
\bibitem{cite_8} J. Cserti, Phys. Rev. B 75, 033405 (2007). 
\bibitem{cite_9} E. J. Nicol and J. P. Carbotte, Phys. Rev. B 77, 155409 (2008).
\bibitem{cite_10} E. McCann, D. S. Abergel, and V. I. Fal’ko, Solid State Commun. 143, 110 (2007). 
\bibitem{cite_11} C. H. Yang, Z. M. Ao,  X. F. Wei and J. J. Jiang, Physica B 457 92 (2015).
\bibitem{cite_12} Wang-Kong Tse and A. H. MacDonald,  Phys. Rev. B 80, 195418 (2009).
\bibitem{cite_13} H Rezania and M Yarmohammadi, Indian J Phys, 90(7), 811 (2016).
\bibitem{cite_14} Palash Nath, D. Sanyal, Debnarayan Jana, Current Applied Physics, 15 691e697 (2015).
\bibitem{cite_15} C.H. Yang, Y. Y. Chen, J. J. Jiang, Z. M. Ao, Solid State Communications 227 23–27 (2016).
\bibitem{cite_16} D. S. L. Abergel, A. Russell, and Vladimir I. Fal’ko, Applied Physics Letters 91, 063125 (2007).  
\bibitem{cite_17} M. Koshino, New Journal of Physics 11 095010 (2009).
\bibitem{cite_18} L.A. Falkovsky Pis’ma v ZhETF, 97 (7), 496-505 (2013) [JETP Lett., 97(7), 429-438 (2013)].
\bibitem{cite_19} L.A. Falkovsky, Jetp, 110(2), 319-324 (2010).
 \bibitem{cite_20} L. M. Zhang, Z. Q. Li, D. N. Basov, M. M. Fogler, Z. Hao and M. C. Martin Phys. Rev. B 78 235408 (2008).
\bibitem{cite_21} You-Chia Chang, Chang-Hua Liu, Che-Hung Liu, Zhaohui Zhong and Theodore B. Norris, Applied Physics Letters, 104, 261909 (2014).
\bibitem{cite_22} M. Bruna and S. Borini, Applied Physics Letters, 94, 031901 (2009).
\bibitem{cite_23} Yingying Wang, Zhenhua Ni, Lei Liu, Yanhong Liu, Chunxiao Cong, Ting Yu, Xiaojun Wang, Dezhen Shen and Zexiang Shen, ACS Nano, 4 (7), pp 4074–4080 (2010).
\bibitem{cite_24} L. A. Falkovsky and S. S. Pershoguba, Phys. Rev. B 76, 153410 (2007).
\bibitem{cite_25} F. Wang, Y. Zhang, C. Tian, C. Girit, A. Zettl, M. Crommie, and Y. R. Shen, Science 320, 206 (2008).
\bibitem{cite_26} B. Huard et al., Phys. Rev. Lett. 98, 236803 (2007).
\bibitem{cite_27} J. R. Williams, L. DiCarlo, C. M. Marcus, Science 317, 638 (2007).
\bibitem{cite_28} D. S. L. Abergel, Hongki Min, E. H. Hwang, and S. Das Sarma, Phys. Rev. B 85, 045411 (2012).
\bibitem{cite_29} V. P. Gusynin, S. G. Sharapov, and J. P. Carbotte, Phys. Rev. Lett. 96, 256802 (2006) 
\bibitem{cite_30} V. P. Gusynin and S. G. Sharapov, Phys. Rev. B 73, 245411 (2006).
\bibitem{cite_31}  Rohit P. Prasankumar, Antoinette J. Taylor (2009). CRC handbook of chemistry and physics:Optical Techniques for Solid-State Materials Characterization. Boca Raton: CRC Press.
\bibitem{cite_32} K. F. Mak, C. H. Lui, J. Shan, and T. F. Heinz, Phys. Rev. Lett. 102,
256405 (2009).
\bibitem{cite_33} Kin Fai Mak, Jie Shan, and Tony F. Heinz, Phys. Rev. Lett. 106,
046401 (2011).
\bibitem{cite_34} V. G. Kravets, A. N. Grigorenko, R. R. Nair, P. Blake, S. Anissimova, K. S. Novoselov, and A. K. Geim, Phys. Rev. B  81, 155413 (2010)
\bibitem{cite_35} Dong-Hun Chae, Tobias Utikal, Siegfried Weisenburger, Harald Giessen, Klaus v. Klitzing, Markus Lippitz, and Jurgen Smet, Nano Lett. 11(3), 1379–1382 (2011).
\bibitem{cite_36} Li Yang, Jack Deslippe, Cheol-Hwan Park, Marvin L. Cohen, and Steven G. Louie, Phys. Rev.Lett. 103, 186802 (2009).
\bibitem{cite_37} Li Yang, Phys. Rev. B, 83, 085405 (2011).
\bibitem{cite_38} Long Ju, et al., Science 358, 907 (2017).
\bibitem{cite_39} Cheol-Hwan Park, and Steven G. Louie, Nano Lett. 10, 426, (2010).
\bibitem{cite_40} Thomas G. Pedersen, Antti-Pekka Jauho and Kjeld Pedersen, Phys. Rev. B 79, 113406 (2009).
\bibitem{cite_41} P. E. Trevisanutto, M. Holzmann, M. C\^{o}t\'e, and V. Olevano, Phys. Rev. B 81, 121405 (2010).
\bibitem{cite_42} V. Apinyan, T.K. Kope\'{c}, Phys. Scr. 91, 095801 (2016).
\bibitem{cite_43} V. Apinyan, T. K. Kope\'{c}, 	Physica E 95, 108 (2018).
\bibitem{cite_44} H. Min, R. Bistritzer, J. J. Su , A. H. MacDonald, Phys. Rev. B 78, 121401 (2008).
\bibitem{cite_45} Y. E. Lozovik, A. A. Sokolik, JETP Lett. 87, 55 (2008).
\bibitem{cite_46} Yu. E. Lozovik, S. L. Ogarkov, A. A. Sokolik , Phys. Rev. B 86, 045429 (2012).
\bibitem{cite_47} Wannier, G., Rev. Mod. Phys. 34, 645 (1962).
\bibitem{cite_48} A. J. Millis, in Strong Interactions in Low Dimensions, edited by D. Baeriswyl and L. De Giorgi (Kluver Academic, Berlin, 2003).
\bibitem{cite_49} A. B. Kuzmenko, I. Crassee, D. van der Marel, P. Blake, and K. S. Novoselov, Phys. Rev. B 80, 165406 (2009).
\bibitem{cite_50} G. D. Mahan, Many-Particle Physics, 3rd ed. (Kluwer Academic/Plenum, New York, 2000).
\bibitem{cite_51} A. A. Abrikosov, L. P. Gorkov, I. E. Dzyaloshinski, Methods of Quantum Field Theory in Statistical Physics, Pergamon Press, (1965).
\bibitem{cite_52} L. A. Falkovsky, J. Phys.: Conf. Ser. 129 012004 (2008).
\bibitem{cite_53} Liu, J.M. Principles of Photonics. Cambridge, United
Kingdom: Cambridge University Press; (2016). 260 p.
DOI: 10.1017/CBO9781316687109.
\bibitem{cite_54} Sosan Cheon, Kenneth David Kihm, Hong goo Kim, Gyumin Lim, Jae Sung Park and Joon Sik Lee, Scientific Reports 4, 6364 (2014).
\bibitem{cite_55} J. P. Hobson and W. A. Nierenberg, Phys. Rev. 89 (1953) 662.
\bibitem{cite_56} M. Abramovitz, I. Stegun, Handbook of Mathematical Functions (Dover, New York, 1970).
\bibitem{cite_57} L. A. Falkovsky, J. Phys.: Conf. Ser. 129 012004 (2008).
\bibitem{cite_58} Sosan Cheon, Kenneth David Kihm, Hong goo Kim, Gyumin Lim, Jae Sung Park and Joon Sik Lee, Scientific Reports 4,  6364 (2014). 





\end{thebibliography}
\end{document}